\documentclass[preprint]{common/sigplanconf}

\usepackage{amsmath}
\usepackage{amsthm}
\usepackage{url}
\usepackage{graphicx}

\usepackage{tikz}
\usepackage{tkz-graph}
\usetikzlibrary{arrows}
\usetikzlibrary{fit}					
\usetikzlibrary{backgrounds}	
\usetikzlibrary{arrows}

\usepackage{color}
\definecolor{Red}{rgb}{0.9,0.0,0.0}  
\definecolor{Green}{rgb}{0.0,0.4,0.0}
\definecolor{Blue}{rgb}{0.0,0.0,0.9}
\definecolor{DarkBlue}{rgb}{0.0,0.0,0.75}
\definecolor{Midnight}{rgb}{0.0,0.0,0.5}
\definecolor{Purple}{rgb}{0.5,0.0,0.4}
\definecolor{Black}{rgb}{0.0,0.0,0.0}
\definecolor{Yellow}{rgb}{1.0,1.0, 0.25}
\definecolor{Cyan}{rgb}{0.25,1.0, 1.0}

\newcommand{\cdColor}{Black}
\newcommand{\kwColor}{DarkBlue}
\newcommand{\comColor}{Red}

\usepackage{listings}
\lstdefinelanguage{SML}{%
  morekeywords={%
    abstype, and, andalso, as, case, datatype, do, else, end, eqtype, exception,%
    fn, fun, functor, handle, if, in, include, infix, infixr, let, local, nonfix,%
    of, op, open, orelse, raise, rec, sharing, sig, signature, struct, structure,%
    then, type, val, where, while, with, withtype,%
  },%
  sensitive,%
  morecomment=[n]{(*}{*)},%
  morestring=[d]",%
}[keywords,comments,strings]%

\lstdefinelanguage{Manticore}[]{SML}{%
  morekeywords={by,otherwise,pcase,pval,spawn,to},
  otherkeywords={(|,|),[|,|],?,\&},
}[keywords,comments,strings]%

\newcommand{\dcfa}{{$\Delta$}{CFA}}
\newcommand{\gcfa}{{$\Gamma$}{CFA}}

\usepackage{algorithm}
\usepackage{algpseudocode}

\usepackage{times}
\SetMathAlphabet{\mathtt}{normal}{OT1}{pcr}{m}{n}
\SetMathAlphabet{\mathtt}{bold}{OT1}{pcr}{bx}{n}

\usepackage{amsmath}
\usepackage{amssymb} 

\newcommand{\CUT}[1]{}

\newcommand{\algoref}[1]{Algorithm~\ref{#1}}

\newcommand{\secref}[1]{Section~\ref{#1}}
\newcommand{\tblref}[1]{Table~\ref{#1}}
\newcommand{\figref}[1]{Figure~\ref{#1}}

\newcommand{\eg}{{\em e.g.}}

\newcommand{\ie}{{\em i.e.}}

\newcommand{\etal}{{\em et al.\/}}

\newcommand{\nesl}{{\textsc{Nesl}}}

\providecommand{\bftt}[1]{{\ttfamily\bfseries{}#1}}

\providecommand{\kw}[1]{\bftt{#1}}
\providecommand{\nt}[1]{{\rmfamily\itshape{#1}}}

\newcount\timeHH
\newcount\timeMM
\timeHH=\time
\divide\timeHH by 60
\timeMM=\time
\multiply\count255 by -60 \advance\timeMM by \count255
\newcommand{\timestamp}{%
  \today{} ---
  \ifnum\timeHH<10 0\fi\number\timeHH\,:\,\ifnum\timeMM<10 0\fi\number\timeMM}

\usepackage{common/code}

\title{Practical Inlining of Functions with Free Variables}

\authorinfo{%
  Lars Bergstrom \\
  John Reppy \\
  Nora Sandler \\
}{%
  University of Chicago}{%
  \{larsberg,jhr,nlsandler\}@cs.uchicago.edu}
\authorinfo{Matthew Fluet}{%
  Rochester Institute of Technology}{%
  mtf@cs.rit.edu}

\date{\today}

\begin{document}

\maketitle
\thispagestyle{empty}

\sloppy

%
\begin{abstract}
A long-standing practical challenge in the optimization of higher-order
languages is inlining functions with free variables.
Inlining code statically at a function call site is safe if the compiler can guarantee that the
free variables have the same bindings at the inlining point as they do at the
point where the function is bound as a closure (code and free variables).
There have been many attempts to create a heuristic to check this correctness
condition, from Shivers' kCFA-based reflow analysis to Might's \dcfa{} and
anodization, but all of those have performance unsuitable for practical compiler
implementations.
In practice, modern language implementations rely on a series of tricks to
capture some common cases (\eg{}, closures whose free variables are only
top-level identifiers such as \lstinline{+}) and rely on hand-inlining by the
programmer for anything more complicated. 

This work provides the first practical, general approach for inlining functions
with free variables.
We also provide a proof of correctness, an evaluation of both the execution time
and performance impact of this optimization, and some tips and tricks for
implementing an efficient and precise control-flow analysis.
\end{abstract}


\section{Introduction}
\label{sec:intro}
Inlining is a program transformation that replaces a function call with the code
body from the target of the call.
This transformation can directly improve performance by eliminating both
function call overheads --- such as stack management, argument passing, and the
jump instruction --- and runtime overheads, such as garbage collection
checks.
Inlining can also indirectly improve performance by optimizing the code body
with respect to actual arguments.
In higher-order languages, function calls can also be made indirectly through a
closure, which is a runtime value that contains a code pointer and an
environment containing the values of bound variables; inlining of first-class functions gains additional potential
performance improvements through removal of the closure value.

Unfortunately, inlining of functions with free variables requires additional
safety analysis.
Optimizing compilers must ensure that the dynamic environment at the static
location where the code can potentially be inlined is the same as the dynamic
environment at all of the closure capture locations that flow to it, up to the
free variables of the function to inline.
This problem is critical, as inlining of functions with free variables comes up
in idiomatic uses of most higher-order languages.
Consider the definition of the \lstinline{map} function in ML, annotated with
a superscript label at an interesting call site:
\begin{lstlisting}[mathescape=true]
val x = 3
fun g i = i + x
fun map f l =
  case l
   of l::ls => (f l)$^1$::(map f ls)
    | _ => []
                 
val res = map g [1,2,3]
\end{lstlisting}

Assuming that there are no other calls to the function \lstinline{map},
we would like to inline the function \lstinline{g} that increments its integer
argument by three into the map function at the call site labeled 1. 
After inlining, constant propagation, and a round of useless variable
elimination, we would obtain the following code:
\begin{lstlisting}[mathescape=true]
fun map l =
  case l
   of l::ls => (l+3)::(map ls)
    | _ => []
                 
val res = map [1,2,3]
\end{lstlisting}

While safe in this specific example, inlining of functions such as
\lstinline{g} is not safe in general because of its free variables,
\lstinline{x} and (less obviously) \lstinline{+}.
The potential lack of safety comes from these free variables which, when
inlined, may have had a different binding at their original capture point than
they have at the inlining point.
Most modern functional compilers make this simple, idiomatic example work
by ignoring any free variables that are bound at the top level when making
inlining decisions.

The following example\footnote{Inspired by one from Might's
  Ph.D. dissertation~\cite{might:environment-analysis}.} exhibits the importance
of reasoning about environments when performing inlining on functions with free
variables.
The function \lstinline{f} takes a boolean and a function of type
\lstinline{unit -> bool}, returning a function of that same type.
If the boolean argument to \lstinline{f} is true, it will return a new function
that simply calls the function it was passed.
If the argument was false, then that parameter is captured in a closure and
returned.
\begin{lstlisting}[mathescape=true]
fun f (b, k) =
  if b then (fn () => (k ())$^1$)
  else (fn () => b)$^2$

val arg = f (false, (fn () => false))
val res = f (true, arg)
val final = res ()
\end{lstlisting}

First, the function \lstinline{f} is called with \lstinline{false}, in order to
capture the variable \lstinline{b} in a closure.
Then, \lstinline{f} is called again, this time with the result of the first
call, to produce a new closure --- in an environment with a different binding
for \lstinline{b} --- that wraps the one from the first call.
Finally, we call that closure.

At the call site labeled 1, control-flow analysis can determine that only
the anonymous function labeled 2 will ever be called.
Unfortunately, if we inline the body of the anonymous function at that
location, as shown in the example code below, the result value \lstinline{final}
will change from \lstinline{false} to \lstinline{true}.
The problem is that the binding of the variable \lstinline{b} is not
the same at the potential inline location as it was at its original capture
location.
\begin{lstlisting}[mathescape=true]
fun f (b, k) =
  if b then (fn () => ((fn () => b) ())$^1$)
  else (fn () => b)$^2$

val arg = f (false, (fn () => false))
val res = f (true, arg)
val final = res()
\end{lstlisting}

While this example is obviously contrived, this problem occurs regularly and the
inability to handle the environment problem in general severely limits most
compilers.

This final example shows a slightly more complicated program that defeats
simple heuristics but in which inlining is safe and the techniques presented in
this work enable.
\begin{lstlisting}[mathescape=true]
let
  val y = m ()
  fun f _ = y
  fun g h = (h ())$^1$
in
  g f
end
\end{lstlisting}
At the call site labeled 1, it is clearly safe to inline the body of the
function \lstinline{f}, since \lstinline{y} has the same binding at the inline
location as the capture location.
Since it is not a trivial idiomatic example, however, it is not commonly
handled.

This work shows how to use control-flow analysis (CFA) to handle many more
inlining situations than are possible in either type-directed optimizers or
simpler heuristics-based optimizers.
By reformulating this \emph{environmental consonance} problem as a graph reachability problem instead of a
partial flow analysis, we have reduced this analysis to a single linear-log time
analysis, independent of the number of potential inlining sites.
This reformulation makes higher-order inlining in the presence of free variables
practical, results in better code performance, and is now on by default in the
Manticore system.
Our contributions are:
\begin{itemize}
 \item A practical heuristic for inlining functions with free variables.
 \item Proof of correctness of that heuristic.
 \item Timing results for both the cost of whole-program higher-order inlining
   analysis ($<$ 3\% of compilation time) and its impact on program performance
   (up to 8\% speedup).
 \item An additional optimization, branch elimination, that is both effective
   and nearly cost-free when already performing control-flow analysis.
\end{itemize}

\paragraph{Roadmap}
First, in the next section we provide a brief introduction to control-flow
analysis and present novel information about tuning the analysis with
respect to recursive datatypes.
\secref{sec:manticore} provides an overview of the Manticore compiler, with a
focus on the continuation-passing style intermediate representation of the
compiler, on which this work is based.
The following sections discuss some of the basic optimizations performed using
the results of control-flow analysis and introduce a straightforward
optimization --- branch elimination --- that takes advantage of a small
extension to the control-flow analysis.
\secref{sec:environment} discusses the analysis challenges presented by
environments in some more detail, and is followed by a section that describes
our novel, \emph{practical} solution.
Finally, we provide empirical data and conclude.

Source code for our complete implementation and all the benchmarks described in this
paper is available at:
\url{http://smlnj-gforge.cs.uchicago.edu/projects/manticore/}.


%
\section{Practical Control-Flow Analysis}
\label{sec:cfa}

This section provides an overview of our specific implementation of control-flow
analysis, placing it in context with other implementations, both theoretical and
practical. 
For a more general introduction to control-flow analysis, in particular the 0CFA
style that we use, the book by Nielson \etal{} provides a comprehensive
introduction to both static program analysis and
0CFA~\cite{principles-prog-analysis}.
While many others have implemented control-flow analysis in their
compilers~\cite{serrano:cfa-paradigm,mlton-cfa,sub-zero-cfa}, our analysis is
novel in its tracking of a wider range of values, including datatypes, and its
lattice coarsening to balance performance and precision.

\subsection{General algorithm}
Using the terminology of Midtgaard's comprehensive survey of control-flow
analysis~\cite{midtgaard-cfa-survey}, our implementation is a
\emph{zeroth-order control-flow analysis} (0CFA).
A 0CFA computes a finite map from all of the variables in a program to a
conservative abstraction of the values that they can take on during the
execution of the code.
Ours does this by starting from an empty map and iterating over the intermediate
representation of the program, merging value flow information into the map based
on the expressions until the map no longer changes.
In our experience, the key to keeping performance acceptable while still
maintaining high precision lies in carefully choosing (and empirically tuning)
the tracked abstraction of values.

\subsection{Tuning the lattice}
Each time we evaluate an expression whose result is bound to a variable, we need
to update the map with a new abstract value that is the result of merging the
old abstract value and the new value given by the analysis.
In theory, if all that we care about in the analysis is the mapping of call
sites to function identifiers, we could use a straightforward domain for the
value map ($\mathcal{V}$), based on the powerset of the function identifiers:
\begin{displaymath}
 \mathcal{V} : \mathtt{VarID} \mapsto 2^{\mathtt{FunID}}
\end{displaymath}
Unfortunately, this domain is insufficiently precise because of the presence of
tuples and datatypes as well as the default Standard~ML calling convention, in
which all arguments are passed as a single argument that is a tuple containing
those arguments.
Without tracking more complicated data structures, we will only be able to track
values in trivial code without datatypes or multi-argument functions.

The representation we use for abstract values therefore is a recursive datatype,
as shown in \figref{fig:values}.
The special $\top$ (\lstinline{TOP}) and $\bot$ (\lstinline{BOT}) elements indicate either all possible values or no
known values, respectively. 
A \lstinline{TUPLE} value handles both the cases of tuples and ML datatype
representations, which by this point in the compiler have been desugared into
either raw values or tagged tuples.
The \lstinline{LAMBDAS} value is used for a set of variable identifiers, all of
which are guaranteed to be function identifiers.
\begin{figure}[ht]
\begin{lstlisting}
  datatype value
    = TOP
    | TUPLE of value list
    | LAMBDAS of CPS.Var.Set.set
    | BOT
\end{lstlisting}
\caption{
  Abstract values.
}
\label{fig:values} 
\end{figure}%

A lattice over these abstract values uses the $\top$ and $\bot$ elements as
usual, and treats values of \lstinline{TUPLE} and \lstinline{LAMBDAS} type as
incomparable.
When two \lstinline{LAMBDAS} values are compared, though, the subset
relationship provides an ordering.
It is this ordering that allows us to incrementally merge flow information, up
to a fixed limit.
The most interesting portion of our implementation is in our handling of merging
two \lstinline{TUPLE} values.
In the trivial recursive solution, the analysis will fail to terminate, due to
the presence of recursive datatypes (\eg{}, on each iteration over a function
that calls the \lstinline{cons} function, we will wrap another \lstinline{TUPLE}
value around the previous value).
In practice for typical Standard ML programs, we have found that limiting the
tracked depth to 5 and then failing any further additions to the $\top$ value
results in good balance of performance and precision.

Note that unlike some other analyses, such as sub-zero CFA, we do not limit the
maximum number of tracked functions per variable~\cite{sub-zero-cfa}.
If the only use of our CFA were inlining, then we could also limit our CFA to a
single potential function; however, as we discuss further in \secref{sec:basic},
this implementation of CFA is used for several optimizations that can still be
performed when multiple functions flow to the same call site.
Further, we did not notice any change in the runtime of the analysis when
reducing the number of tracked function variables, but it severely impacts our
ability to perform some of those other optimizations.

\subsection{Adding booleans for additional flow-sensitivity}
\label{sec:cfa:bool}
Normally, CFA does not track booleans and other raw values, so they are
represented by the $\top$ value.
To track booleans, we add \lstinline{true} and \lstinline{false} to our list of abstract values,
and update CFA to account for their place in the lattice.
When both \lstinline{true} and \lstinline{false} values flow to the same
variable, we promote its abstract value to $\top$.
The updated datatype is shown in \figref{fig:values-bool}.
\begin{figure}[ht]
\begin{lstlisting}
  datatype value
    = TOP
    | TUPLE of value list
    | LAMBDAS of CPS.Var.Set.set
    | BOOL of bool
    | BOT
\end{lstlisting}
\caption{
  Abstract values with boolean tracking.
}
\label{fig:values-bool} 
\end{figure}%


%
\section{Manticore}
\label{sec:manticore}
In order to provide more details on both the implementation of control-flow
analysis and the optimizations that rely on it, we provide some background on
the host compiler and relevant intermediate representation.
The compiler operates on the whole program at once, reading in the files in the
source code alongside the sources from the runtime library.
As covered in more detail in an earlier paper~\cite{manticore-ml07}, there are
six distinct intermediate representations (IRs) in the Manticore compiler:
\begin{enumerate}
  \item Parse tree --- the product of the parser.
  \item AST --- an explicitly-typed abstract-syntax tree representation.
  \item BOM --- a direct-style normalized
    $\lambda$-calculus.
  \item CPS --- a continuation passing style $\lambda$-calculus.
  \item CFG --- a first-order control-flow-graph representation.
  \item MLTree --- the expression tree representation used by the
    MLRISC code generation framework~\cite{mlrisc}.
\end{enumerate}%
The work in this paper is performed on only the CPS representation.

\subsection{CPS}
Continuation passing style (CPS) is the final high-level representation used in
the compiler before closure conversion generates a first-order representation
suitable for code generation.
CPS transformation is performed in the Danvy-Filinski
style~\cite{representing-control:cps-study}.
This representation is a good fit for a simple implementation of control-flow
analysis because it transforms each function return into a call to another
function.
The uniformity of treating all control-flow as function invocations simplifies
the implementation.
We also have an implementation of control-flow analysis on the BOM direct-style
representation, which was used for some Concurrent ML-specific
optimizations~\cite{cml-specialization}.
The BOM-based implementation is almost 10\% larger in lines of code, despite
lacking the optional features, user-visible controls, and optimizations
described in this paper.

The primary datatypes and their constructors are shown in \figref{fig:cps-ir}.
Key features of this representation are:
\begin{itemize}
  \item Each expression has a program point associated with it, which serves as
    a unique label.
  \item It has been normalized so that every expression is bound to a variable.
  \item The \lstinline{rhs} datatype, not shown here, contains only immediate
    primitive operations.
  \item The CPS constraint is captured in the IR itself --- \lstinline{Apply}
    and \lstinline{Throw} are non-recursive constructors, and there is no way to
    sequence an operation after them.
\end{itemize}%

\begin{figure}[ht]
\begin{lstlisting}[mathescape=true]
datatype exp = Exp of (ProgPt.ppt * term)
and term
  = Let of (var list * rhs * exp)
  | Fun of (lambda list * exp)
  | Cont of (lambda * exp)
  | If of (cond * exp * exp)
  | Switch of (var * (tag * exp) list * exp option)
  | Apply of (var * var list * var list)
  | Throw of (var * var list)
and lambda = FB of {
  f : var,
  params : var list,
  rets : var list,
  body : exp
}
and $\dots$
\end{lstlisting}%
  \caption{Manticore CPS intermediate representation.}
  \label{fig:cps-ir}
\end{figure}%


%
\section{Basic Optimizations}
\label{sec:basic}
In Manticore, we use the data from control-flow analysis (CFA) to augment some
of our optimizations.
In this section, we list some of those optimizations and describe the
differences between purely type-directed optimizations and those that rely on
the results of CFA. 

\subsection{Argument flattening}
The default calling convention in Standard ML (and many other functional
languages) involves heap-allocating all arguments to a function and passing a
pointer to that heap-allocated data to the function.
As has been described in the Haskell literature, we could analyze the type of
the function and then, based on the calling convention of that type, place the
arguments in appropriate registers and evaluate the
function~\cite{haskell-apply}.
In many cases, when combined with inlining and other optimizations, we can then
avoid allocating some arguments in the heap.
By instead using CFA as the basis for our optimizations, we can not only
specialize the calling convention but have several additional optimization
opportunities:
\begin{enumerate}
  \item If the target function or functions only use a few of the parameters, we
    can change their calling convention and type to only pass those parameters.
  \item We adjust both callers and callees to flatten tuples in the case where
    the caller performs the allocation and the callee simply extracts members of
    that tuple.
\end{enumerate}
This work has been described in detail in an earlier
paper~\cite{arity-raising}.

\subsection{Calling conventions}
There are two interesting optimizations related to CFA in the Manticore
implementation of calling conventions.
In the following example, even in a type-directed optimizer, inlining will cause
\lstinline{g} to be inlined and thus the call to \lstinline{f} will become known and eligible
for a direct jump:
\begin{lstlisting}
let 
  fun f x = x + 1
  fun g (h, i) = h i
in
  g (f, 3)
end
\end{lstlisting}
But, if \lstinline{g} is too large to inline and if there are any other functions of
the same type as \lstinline{f}, it will be unclear to the compiler what functions could
be bound to the variable \lstinline{h}, forcing it to make an indirect call through a
pointer to invoke \lstinline{f}.
In Manticore, we use CFA to recognize that \lstinline{f} is known and perform a
direct jump to it within the body of \lstinline{g}.

Even in the case where \lstinline{f} has free variables, and thus we need to pass a
closure, an additional optimization is still available.
Typically, closures include both a code pointer and the associated environment
data.
In this case, though, we replace the closure with just the environment data and
instead perform a direct jump.
Both of these optimizations remove just a single pointer indirection and a few
bytes of allocation, but the availability of these optimizations means that
users of the system do not have to transform their code by hand to get good
performance in key inner loops.



\section{Branch elimination}
\label{sec:elimination}
In \secref{sec:cfa:bool}, we showed an extension of the lattice of abstract values
computed by control-flow analysis that also tracks boolean values.
In this section, we describe a compiler optimization that eliminates conditional
branches that will never be taken based on these values.
Consider the following function:
\begin{lstlisting}
fun f (boolean) =
  if boolean
  then g 3
  else g 4
\end{lstlisting}
If the value of \lstinline{boolean} is known, we can eliminate the conditional statement 
and leave only the relevant branch.

Early in the compilation process of Manticore, between the AST and BOM phases
described in \secref{sec:manticore}, the \lstinline{if} construct on a boolean
is converted into a case over booleans with constant values.
The example above is therefore translated into roughly the following code, in
direct-style instead of continuation-passing style for consistency with
the rest of the examples in this paper:
\begin{lstlisting}
fun f (boolean) =
 case boolean
   of 0x0 => g 4
    | 0x1 => g 3
\end{lstlisting}

Suppose this function occurs in a larger program that only calls \lstinline{f}
with arguments whose values are \lstinline{true}.
In that case, the control-flow analysis will have an abstract value of
\lstinline{BOOL(true)} associated with the variable \lstinline{boolean}.
Then, the branch elimination pass can eliminate all but the arm of the
\lstinline{case} expression that corresponds to the \lstinline{true} case.
After that branch elimination and a useless variable elimination, we are left
with the following program:
\begin{lstlisting}
fun f () = g 3
\end{lstlisting}

Since this optimization relies on information already computed by the
control-flow analysis and requires only a single pass over the intermediate
representation, as shown in \secref{sec:evaluation}, it has nearly zero cost.


%
\section{Environment problems}
\label{sec:environment}
Some optimizations that are straightforward in a first-order language when
combined with control-flow analysis are not safe in a higher-order language
using only the results of a zeroth-order CFA as described in \secref{sec:cfa}.
In that description of the CFA, the abstraction of the environment is a single,
global map for each variable to a single value from the lattice.
This restriction means that the CFA results alone do not allow us to reason
separately about bindings to the same variable that occur along different
control-flow paths of the program.

This restriction impedes inlining.
Inlining a function is a semantically safe operation when a call site is
to a unique function and that function has no free variables.
But, in a higher-order language, inlining of a function with free variables is
only safe when those free variables are guaranteed to have the same
bound value at the capture location and inlining location, a property which
Shivers called \emph{environmental consonance}~\cite{shivers:phd-thesis}.
For example, in the following code, if CFA determines that the function \lstinline{g}
is the only one ever bound to \lstinline{f}, then the body of \lstinline{g} may
be inlined at the call site labeled 1.
\begin{lstlisting}[mathescape=true]
val x = 3
fun g i = i + x
fun map f l =
  case l
   of l::ls => (f l)$^1$::(map f ls)
    | _ => []
                 
val res = map g [1,2,3]
\end{lstlisting}

While many compilers special-case this particular situation, in which all free variables of the function are bound at the top level, even small changes
break these fragile optimizers, as shown in the following code:
\begin{lstlisting}
fun wrapper x = let
  fun g i = i + x
  fun map f l =
    case l
     of l::ls => (f l)$^1$::(map f ls)
      | _ => []
in
  map g [1,2,3]
end
val res1 = wrapper 1
val res2 = wrapper 2
\end{lstlisting}
Performing the inlining operation is again safe, but the analysis required to
guarantee that the value of \lstinline{x} is always the same at both the body of the
function \lstinline{wrapper} and in the call location inside of \lstinline{map} is beyond
simple heuristics.

\paragraph{Copy propagation}
A similar operation that has the identical problem is copy propagation.
In this operation, instead of inlining the body of the function (\eg{}, because
it is too large), we are attempting to remove the creation of a closure by
turning an indirect call through a variable into a direct call to a target
function.
In the following code, the function \lstinline{g} is passed as an argument to \lstinline{map}
and called in its body.
\pagebreak
\begin{lstlisting}
fun map f l =
  case l
   of l::ls => (f l)::(map f ls)
    | _ => []

val res = map g [1,2,3]
\end{lstlisting}
When \lstinline{g} either has no free variables or we know that those free variables
will always have the same values at both the capture and inlining location, we
can substitute \lstinline{g}, potentially removing a closure and enabling the compiler
to optimize the call into a direct jump instead of an indirect jump through the
function pointer stored in the closure record.

\paragraph{Interactions}
These optimizations are not only important because they remove an indirect call.
If the variable that previously held the closure is no longer used, then the
useless variable elimination pass will remove it from all calls and as a
parameter to the function, as shown in the code below, reducing register
pressure.
\begin{lstlisting}
fun map l =
  case l
   of l::ls => (g l)::(map ls)
    | _ => []

val res = map [1,2,3]
\end{lstlisting}


%
\section{Reflow}
\label{sec:reflow}
A theoretical solution to this environment problem that enables a suite of
additional optimizations is reflow analysis~\cite{shivers:phd-thesis}.
Traditional reflow analysis requires re-running control-flow analysis from the
potential inlining point and seeing if the variable bindings for all relevant
free variables are uniquely bound with respect to that sub-flow.
Unfortunately, this operation is potentially quite expensive (up to the same
complexity as the original CFA, at each potential inlining site) and no compiler
performs it in practice.

We use a novel analysis that builds upon the static control-flow graph of the
program.
Our goal was to build up a data structure that could perform --- in less than
linear time --- the same test used in reflow analysis.
The optimizations from \secref{sec:environment} are safe when the free variables
of the target function are guaranteed to be the same at its closure
creation point and at the target call site.
In the original work by Shivers, this question was answered by checking whether
a binding for a variable had changed between those two locations via a
re-execution of control-flow analysis.
Our analysis instead turns that question into one of graph reachability: in the
graph corresponding to the possible executions of this program, is there any
possible path between those two locations through a rebinding of the free
variables?

This graph is built in two steps.
First, build a static control-flow graph for each function, ignoring function
calls, annotated with variable bindings and rebindings.
Then, augment those individual function graphs with edges from the call sites 
to the potential target functions, as determined by the control-flow analysis.
Though we discuss only our implementation of 0CFA in this work, this approach to
reflow analysis also works with other control-flow analyses.

The variable bindings and rebindings in a program written in the
continuation-passing style (CPS) representation defined in \figref{fig:cps-ir}
happen in two cases:
\begin{itemize}
  \item At the definition of the variable, which is either a \lstinline{let}-binding or
    as a parameter of a function.
  \item In the case when a free variable of a function was captured in a
    closure and this captured value is restored for the execution of that
    function.
\end{itemize}
We capture both of these conditions through labeled nodes in the graph for
each function.
One node is labeled with all of the free variables of the function,
since those are the ones that will be rebound when the function is called
through a closure.
A second node is labeled with all of the parameters to the function, since they
will also be bound when the function is called.
Finally, any \lstinline{let}-binding in the control-flow graph will be labeled
with the variable being bound.

Augmentation of the call sites is done using the results of the control-flow
analysis described in \secref{sec:cfa}.
In the intermediate representation, all targets of call sites are variables.
In the trivial case, that variable is the name of a function identifier, and we
can simply add an edge from the call site to that function's entry point.
Otherwise, that variable is of function type but can be bound to many possible
functions.
In that case, the control-flow analysis will provide one of three results:
\begin{itemize}
  \item The value $\bot$, indicating that the call site can never be reached in
    any program execution.
    No changes are made to the program graph in this case.
  \item The value $\top$, indicating that any call site may be reached.
    In this case, we add an edge to a special node that represents any call
    site, whose optimization is discussed in \secref{sec:reflow:imprecision}.
  \item A set of function identifiers.
    Here, we add one edge from the call site per function, to that function's
    entry point.
\end{itemize}

At this point, the graph is complete and enables us to reformulate the safety
property.
We can now simply ask: does there exist a path between the closure capture
location and the target call site in the graph that passes through a rebinding
location for the free variables of the function that we want to inline?

\subsection{Safe example}
\begin{figure}[ht]
  \begin{center}
    \includegraphics[scale=0.75]{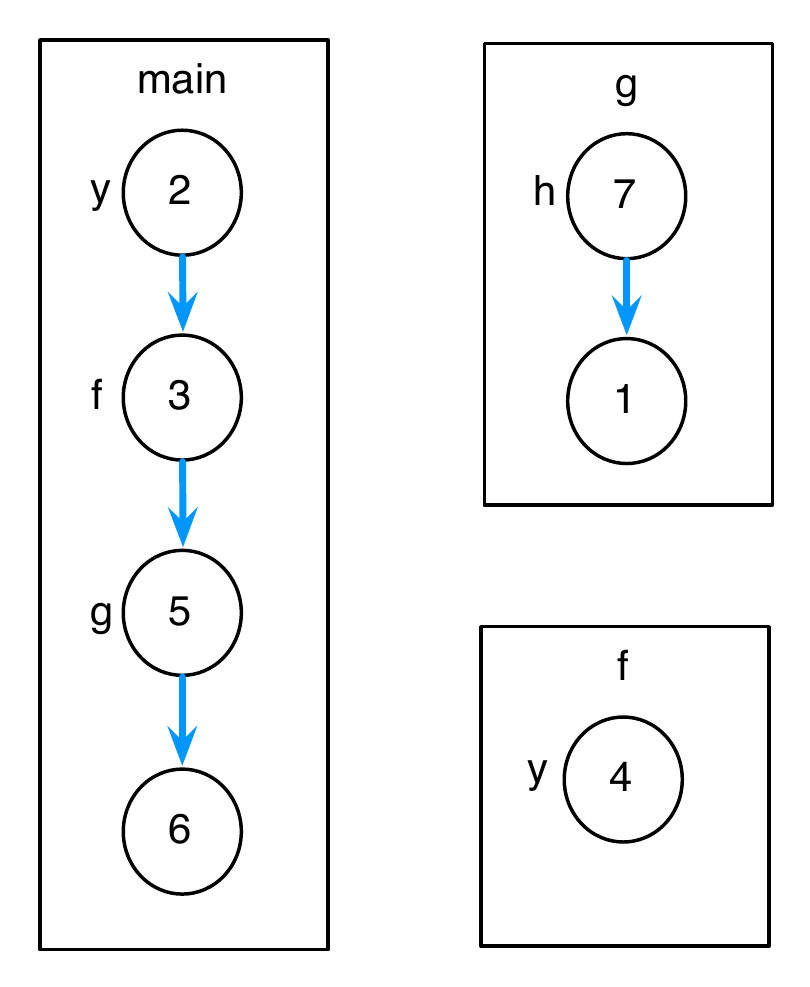}
  \end{center}%
  \caption{Control-flow graph for a safe example, before adding edges for call
    sites.}
  \label{fig:example2a}
\end{figure}

We first revisit the first complicated, but safe, example from the introduction,
annotated with additional labels for use in the graph:
\begin{lstlisting}[mathescape=true]
let
  val y$^2$ = 2
  fun f$^3$ _ = y$^4$
  fun g$^5$ h$^7$ = (h ())$^1$
in
  (g f)$^6$
end
\end{lstlisting}
In the first stage of building the graph, we create the static control-flow
graph for each function.
That graph is shown in \figref{fig:example2a}.
Nodes in this graph represent the labeled expressions from the source program,
and edges are the static control-flow.
We also create a separate function for toplevel bindings and the program's
control-flow from the entry point.

\begin{figure}[ht]
  \begin{center}
    \includegraphics[scale=0.75]{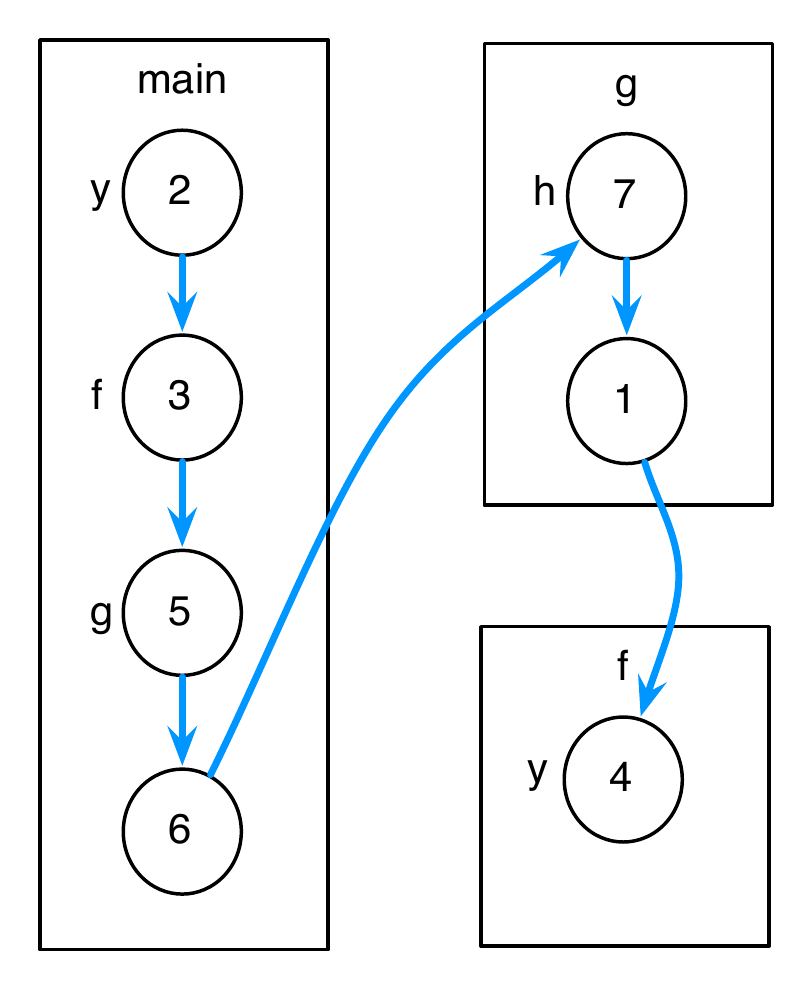}
  \end{center}%
  \caption{Control-flow graph for a safe example, with edges for call sites.}
  \label{fig:example2b}
\end{figure}

Then, we augment that graph with edges from each of the call sites to the target
functions, based on the results of CFA.
That graph appears in \figref{fig:example2b}.

Revisiting the question of whether it is safe to inline the body of the function
\lstinline{f} at the call site labeled 1, we now turn it into a graph question.
Does there exist a path from the closure capture location (node 5), through either of
the rebinding sites for the free variable \lstinline{y} (nodes 2 and 4), that
terminates at the potential inlining site (node 1)?
Since one does not exist in this graph, the inlining is safe to perform.

\subsection{Unsafe example}
\begin{figure}[ht]
  \begin{center}
    \includegraphics[scale=0.75]{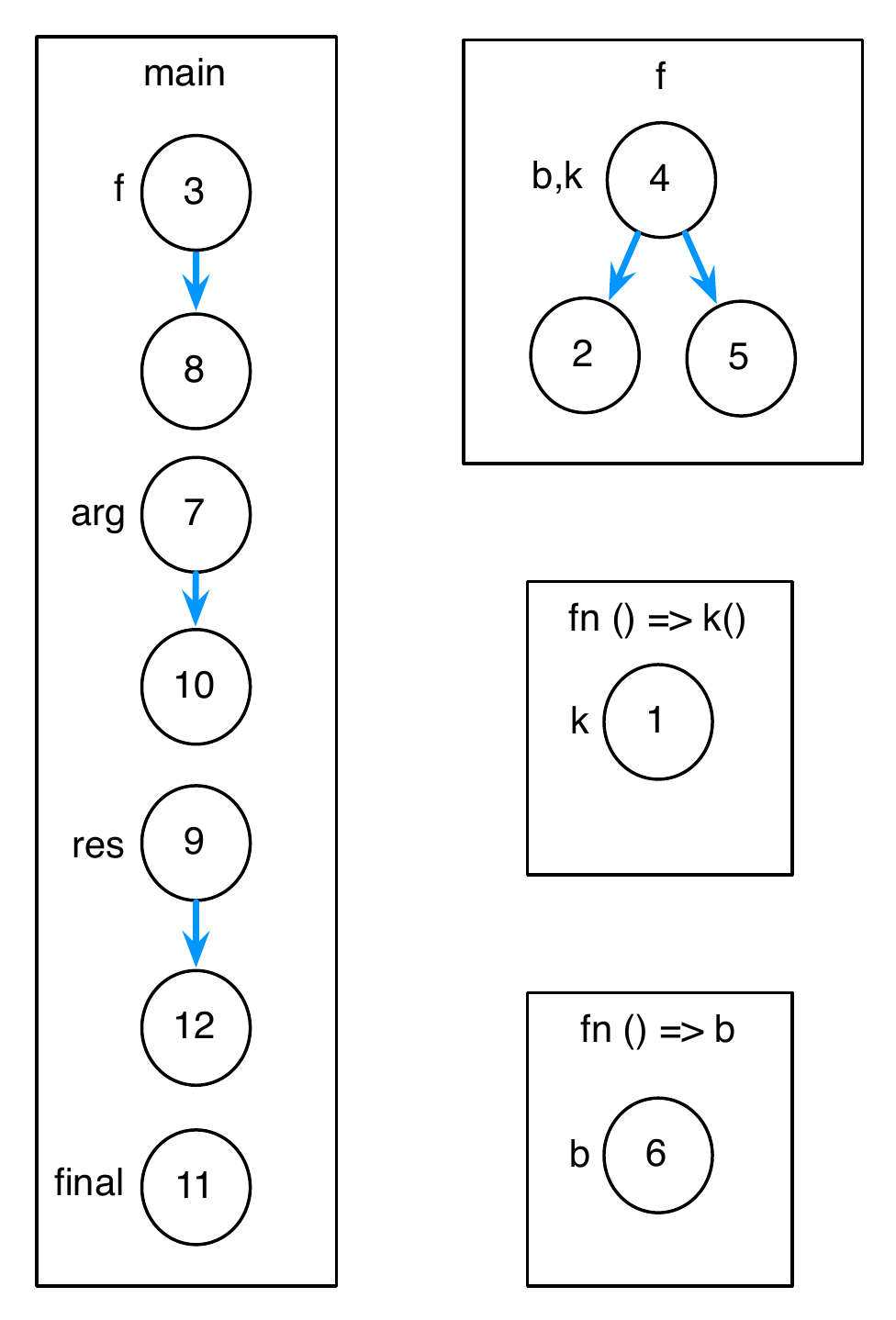}
  \end{center}%
  \caption{Control-flow graph for an unsafe example, before adding edges for call
    sites.}
  \label{fig:example3a}
\end{figure}

For a negative case, we revisit the unsafe example from the introduction, with
additional labeling added.
\pagebreak
\begin{lstlisting}[mathescape=true]
fun f$^3$ (b, k)$^4$ =
  if b then (fn () => (k ())$^1$)$^5$
  else (fn () => b$^6$)$^2$

val arg$^7$ = (f (false, (fn () => false)))$^8$
val res$^9$ = (f (true, arg))$^{10}$
val final$^{11}$ = (res ())$^{12}$
\end{lstlisting}
Again, first we build a graph of the static control-flow for each function,
shown in \figref{fig:example3a}.

\begin{figure}[htb]
  \begin{center}
    \includegraphics[scale=0.75]{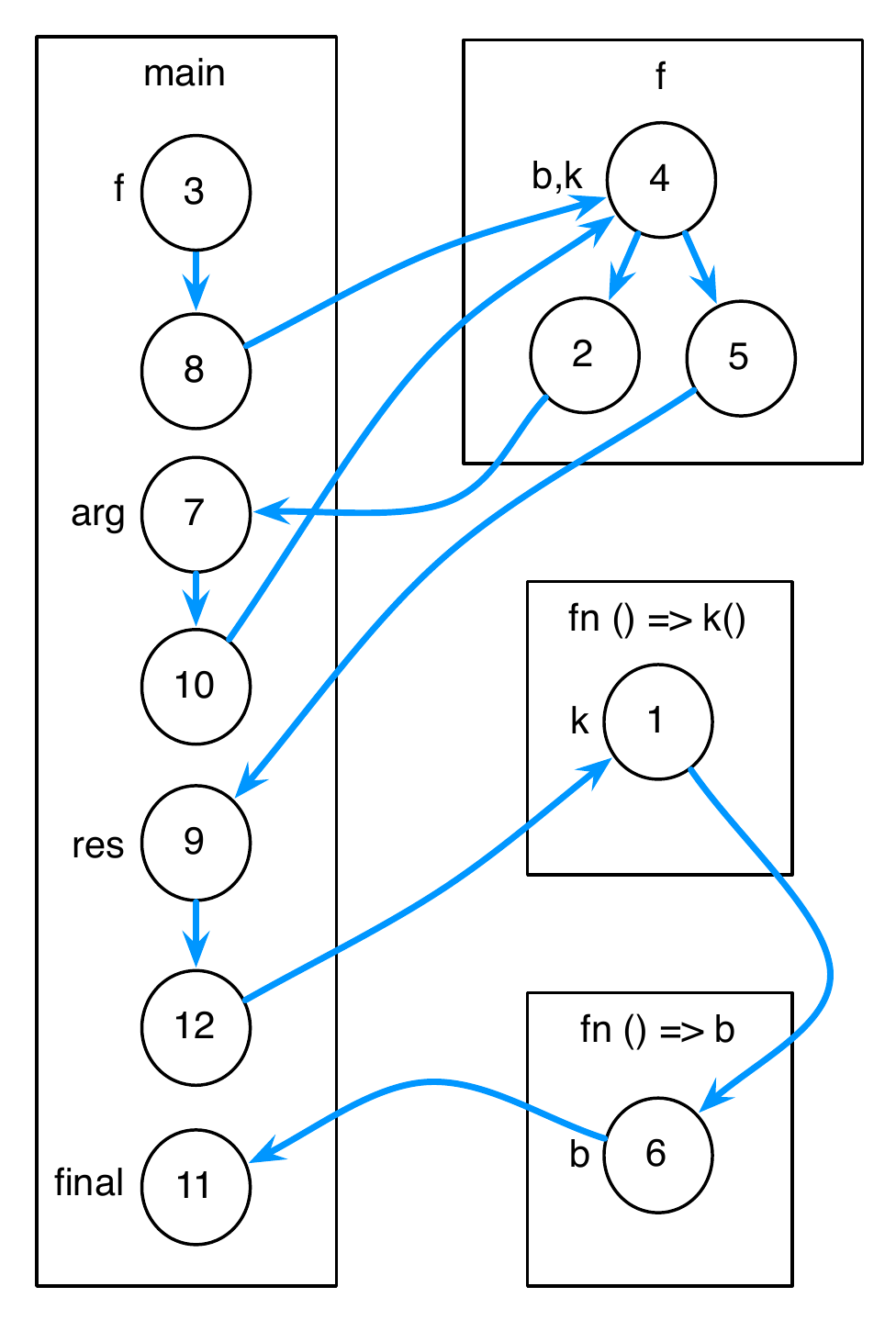}
  \end{center}%
  \caption{Control-flow graph for an unsafe example, with edges for call sites.}
  \label{fig:example3b}
\end{figure}

Then, we augment that graph with edges from each of the call sites to the target
functions, based on the results of CFA.
That graph appears in \figref{fig:example3b}.
The nodes labeled 2, 5, and 6 are of particular interest.
In the direct-style code we have used in this example, the outgoing edge from
each of them corresponds to the return point of the function.
In the continuation-passing style IR used in the compiler for this analysis,
those have been translated into calls to functions that correspond to the return
point.

In this example, we were investigating whether it was safe to inline the
anonymous function defined at label 2 at the call site labeled 1.\footnote{
  Knowing that $V($\lstinline{k}$)$ = $\{$ \lstinline{(fn () => b)} $\}$ instead of a larger
  set requires an analysis more precise than 0CFA, such as our
  RC-CFA~\cite{bergstrom:masters-paper} or Might's \gcfa{}~\cite{might:gamma}.
}
So, in the graph in \figref{fig:example3b}, does there exist a path from the
closure capture location (2) to the potential inlining point (1) that passes
through a binding location for \lstinline{b} (4 or 6)?
Since we can find such a path in the graph (\eg{},
$2\to7\to10\to4\to5\to9\to12\to1$), this inlining is potentially (and actually)
unsafe, so it is disallowed under the reflow condition tested in our system.

\subsection{Computing graph reachability quickly}
\label{sec:reachability}
This question about the existence of paths between nodes in the graph is a
reachability problem.
There are off-the-shelf $O(n^3)$ algorithms such as Warshall's algorithm for
computing graph reachability~\cite{warshall}, but those are far too slow for
practical use. 
On even small graphs of thousands of nodes, they take seconds to run.
Other optimized algorithms have runtimes on the order of $O(n*\mathit{log}\ n)$,
but this is the runtime for each query and we might perform up to $O(n)$ queries
for a particular program.

Therefore, we use an approach that collapses the graph quickly into a map we can
use for logarithmic-time queries of the reachability between two nodes.
The approach we use performs two steps.
First, we take the potentially cyclic graph and reduce it into a set of
strongly-connected components, which takes $O(n*\mathit{log}\ n)$.
Then, we use a bottom-up approach to compute reachability in the resulting DAG,
which is $O(n)$.
All queries are then performed against the resulting map from source component
to set of reachable components.
The membership test within that set takes $O(\mathit{log}\ n)$.
The overall complexity of our algorithm is therefore $O(n*\mathit{log}\ n)$.

\paragraph{Strongly-connected components}
We use Tarjan's algorithm for computing the strongly-connected
components~\cite{tarjan-scc}, as implemented in Standard ML of New Jersey by
Matthias Blume.
This produces a directed acyclic graph (DAG).
There are two interesting types of components for this algorithm: those that
correspond to only one program point and those that correspond to more than one
program point. 
In the single point case, control-flow from that point cannot reach itself.
When there are multiple points, control-flow \emph{can} reach itself.
This distinction is crucial when initializing the reachability map.

\paragraph{Reachability in a DAG}
Starting from the leaf components of the graph, we add those components and
everything that they can reach to the reachability set associated with each of
their parent components.
As those leaves are handled and removed from the worklist, their parents become
leaves and we handle them iteratively until there are no components remaining
in the graph.
A more detailed description is shown in \algoref{alg:dag}.
\begin{algorithm}
\caption{Compute DAG reachability for a graph $\mathit{DAG}$}
\label{alg:dag}
\begin{algorithmic}
\For{$\mathit{node} \in \mathit{DAG}$}
  \State $Done(\mathit{node}) \leftarrow \mathbf{false}$
  \If{$\#\mathit{points} \in \mathit{node} = 1$}
    \State $R(\mathit{node}) \leftarrow \{\}$
  \Else
    \State $R(\mathit{node}) \leftarrow \{\mathit{node}\}$
  \EndIf
\EndFor
\State $\mathit{leaves} \leftarrow$ all leaves in $\mathit{DAG}$
\While{$leaves$ is not empty}
 \State $\mathit{leaf} \leftarrow$ a leaf in $leaves$
 \State $leaves \leftarrow leaves - \mathit{leaf}$
 \State $Done(leaf) \leftarrow \mathbf{true}$
 \For{$p \in parents(\mathit{leaf})$}
   \State $R(p) \leftarrow R(p) \cup R(\mathit{leaf}) \cup \{leaf\}$
   \If{$(\forall{c \in Children(p)})(Done(c) = \mathbf{true})$}
     \State $\mathit{leaves} \leftarrow \mathit{leaves} \cup \{p\}$
   \EndIf
 \EndFor
\EndWhile

\end{algorithmic}
\end{algorithm}

\subsection{Handling imprecision}
\label{sec:reflow:imprecision}
In practical implementation, we also need to handle a variety of sources of
imprecision.
C foreign function calls, the entry and exit point of the generated binary
itself (\ie{}, the \lstinline{main} function), and the limited lattice size all
contribute to situations where a call site may be through a variable whose
target is $\top$, or unknown.
The obvious way to handle this situation when creating the graph is to add an
edge from any call site labeled $\top$ to every possible function entry point.
Unfortunately, that frequently connects the entire graph, preventing the
compiler from inlining any function at all.
Instead, we take advantage of the fact that a call to any function is
really only a call to any function where not all its callers are known.
We therefore add an edge from any call site labeled $\top$ to any function whose
callers are not all known.
Since those functions are a relatively small set, the graph remains useful.

\subsection{Safety}
We prove safety of this algorithm by building on the conditions of
correctness for inlining functions with free variables from Might and Shivers in
their work on \dcfa~\cite{might:delta}.
In that work, they prove that two conditions are sufficient for 
semantics-preserving inlining of a function:
\begin{enumerate}
  \item All closures invoked at the given call site are to the same function.
  \item The environment at the call site is equivalent up to the free variables
    of that function to the environment within any captured closure that reaches
    it.
\end{enumerate}

To show the correctness of our algorithm, we therefore need to show that there
is no case where our algorithm will attempt to perform an inlining operation,
but the two \dcfa{} inlining conditions do not hold.
For contradiction, assume that there exists such an invalid inlining operation
from our algorithm.
Then, there are two cases to handle:

\subsubsection{Same function}
By construction, our 0CFA-based algorithm has more conservative (\ie{}, coarser)
results than the idealized \dcfa{} algorithm.
Therefore, whenever we identify a call site as only invoking closures of a given
function, so must \dcfa.

\subsubsection{Environment equivalence}
For the captured free variables in one of the closures to be different between
its capture point and inlining location, there must have been a change in those
variable bindings between those two locations along a path in the control-flow
graph. 
A variable binding can either have been \emph{superseded} by a newer binding or
\emph{reverted} to an older binding.

We know that the binding cannot have been superseded because we have added all
of the binding locations for the variables in the program to our control-flow
graph and then only allow inlining in the case where there is no path from the
closure capture location to one of these binding locations and then on to the
inlining point.

Reversion to an earlier binding happens when a different binding to a variable
has been captured in a closure which is then restored through the application of
that closure.
However, the control-flow graph for each function is augmented with a program
point annotated with any free variables used in the function. 
These program points are also treated as binding locations, so we allow inlining
only if no path from capture location to inlining location passes through such a
binding location.

Note that we are in a CPS-based representation and function calls never return.
In a direct-style intermediate representation, the control-flow graph would
similarly need rebinding locations for all of the variables that return into
scope at each return point.

\subsection{Limitations}
While safe, this analysis necessarily is more limited than general formulations
of higher-order inlining as shown by Shivers' kCFA framework (for larger values
of k than 0) or Might's \dcfa{} approach~\cite{shivers:phd-thesis, might:delta}.
Both of those analyses are able to distinguish environments created by different
control-flow paths through the program.
Our analysis collapses all different control-flow paths to each function,
resulting in a potential loss of precision.
For example, in the following program, which is an extension of the motivating
example from the introduction, our attempt to inline at the call site labeled 1
will fail.
After the first call to \lstinline{callsG}, the function \lstinline{confounding}
is in the abstract possible set of functions that can be bound to the parameter
\lstinline{k}.
Even though in the first call the boolean tracking avoids analyzing
\lstinline{g} and adding \lstinline{confounding} to the list of possible values
for \lstinline{h}, when the second call comes through, the function
\lstinline{f} is added to the possible set of values for \lstinline{k} and then
\emph{both} of those are added to the set of values that could be bound to
\lstinline{h}.
Fundamentally, this problem is the one that stronger forms of control-flow
analysis handle, though clearly there are some heuristics that could be used to
increase the precision on this specific case.
\begin{lstlisting}[mathescape=true]
let val y = 2
  fun f _ = y
  fun confounding _ = raise Fail ""
  fun g h = h$^1$()
  fun callsG b k = if b then g k else 0
  val bad = callsG false confounding
in
  callsG true f
end
\end{lstlisting}


%
\section{Evaluation}
\label{sec:evaluation}

\subsection{Experimental method}
Our benchmark machine has two 8~core Intel Xeon E5-2687 processors running at
3.10~GHz.
It has 64~GB of physical memory.
This machine runs x86\_64 Ubuntu Linux 11.10, kernel version 3.0.0-30.

We ran each benchmark experiment 30 times, and we report the median runtime and
standard deviation in our tables.
Times are reported in seconds.

This work has been implemented, tested, and is part of the current Manticore
compiler's default optimization suite.
In addition to inlining, we also perform copy propagation, which replaces a
variable that calls a function through a closure with the name of the function
itself and requires an identical safety condition.

\subsection{Benchmarks}
For our empirical evaluation, we use six benchmark programs from our parallel
benchmark suite. 
Each benchmark is written in a pure, functional style.

The Barnes-Hut benchmark~\cite{barnes-hut} is a classic N-body problem solver.
Each iteration has two phases.
In the first phase, a quadtree is constructed from a sequence of mass points.
The second phase then uses this tree to accelerate the computation of
the gravitational force on the bodies in the system.
Our benchmark runs 20 iterations over 400,000 particles generated in
a random Plummer distribution.
Our version is a translation of a Haskell program~\cite{barnes-hut-haskell-bench}.

The Raytracer benchmark renders a $2048 \times 2048$ image as two-dimensional
sequence, which is then written to a file.
The original program was written in ID~\cite{id90-manual} and is a simple
ray tracer that does not use any acceleration data structures.

The Mandelbrot benchmark computes the Mandelbrot set, writing its output to an
image file of size $2048 \times 2048$.

The Quickhull benchmark determines the convex hull of 8,000,000 points in the
plane.
Our code is based on the algorithm by Barber \etal{}~\cite{quickhull}.

The Quicksort benchmark sorts a sequence of 5,000,000 integers in parallel.
This code is based on the \nesl{} version of the algorithm~\cite{scandal-algorithms}.

The SMVM benchmark is a sparse-matrix by dense-vector multiplication.
The matrix contains 3,005,788 elements, the vector contains 10,000, and the
multiplication is iterated 25~times. 

\begin{table*}[ht]
\begin{center}
\begin{tabular}{r || r | r | r | r | r | r | r | r}
 &  &  &  &  & \multicolumn{1}{c|}{Branch} & \multicolumn{1}{c|}{Copy} & &  \\
Benchmark & Lines & Expressions & Total (s) & CFA (s) & Elim. (s) & Prop. (s) & Inline (s) & GCC (s)  \\
\hline
Barnes-hut & 334 & 17,400 & 8.79 & 0.042 & 0.003 & 0.175 & 0.198 & 2.56  \\ 
Raytracer & 501 & 12,800 & 6.54 & 0.019 & 0.002 & 0.112 & 0.124 & 2.64 \\
Mandelbrot & 85 & 9,900 & 5.06 & 0.013 & 0.006 & 0.091 & 0.098 & 1.70 \\
Quickhull & 196 & 15,200 & 7.67 & 0.039 & 0.003 & 0.182 & 0.177 & 2.05 \\
Quicksort & 74 & 11,900 & 5.49  & 0.022 & 0.001 & 0.111 & 0.122 & 1.11 \\
SMVM & 106 & 13,900 & 7.25 & 0.033 & 0.002 & 0.131 & 0.123 & 2.52 \\
\end{tabular}%
\end{center}
\caption{
  Benchmark program sizes, both in source lines and total number of expressions in 
  our whole-program compilation.
  Costs of the analyses and optimizations are also provided, in seconds.
}
\label{eval:benchmarks}
\end{table*}%

\subsection{Compilation performance}
In \tblref{eval:benchmarks}, we have broken down the compilation time of these
parallel benchmarks.
While we have included the number of lines of code of the benchmarks, Manticore
is a whole-program compiler, including the entire basis library.
Therefore, in addition to the lines of code, we have also reported the number of
expressions, where an expression is an individual term from the intermediate
representation shown in \figref{fig:cps-ir}.
By that stage in the compilation process, all unreferenced and dead code has
been removed from the program.

The most important results are:
\begin{itemize}
  \item Control-flow analysis and branch elimination are basically free.
  \item The reflow analysis presented in this work (which represents the
    majority of the time spent in both the copy propagation and inlining passes)
    generally makes up 1-2\% of the overall compilation time.
  \item Time spent in the C compiler, GCC, generating final object code is the
    longest single stage in our compiler.
\end{itemize}

\subsection{Benchmark performance}
Across our already tuned benchmark suite, we see several improvements and only
one statistically significant slowdown, as shown in
\tblref{benchmarks:optimization}.
The largest challenge with analyzing the results of this work is that for any
tuned benchmark suite, the implementers will have already analyzed and removed
most opportunities for improvement.
When we investigated the usefulness of these optimizations on some programs we
ported from a very highly tuned benchmark suite, the Computer Language Benchmark
Game~\cite{shootout}, we could find zero opportunities for further
optimization.

All three of the major optimizations described in this paper --- branch
elimination, copy propagation, and higher-order inlining --- are applied to the
optimized version, and none of them are present in the baseline.
We run copy propagation before inlining in the compiler because that enables us
to re-use the control-flow analysis information. If we performed inlining first,
we would have to run CFA an additional time or manually update that information
during the transformations before we could perform copy propagation.

Mandelbrot especially benefits from these optimizations, with a 12\% improvement
on one processor and 7.8\% improvement in parallel.
In this program, there is only one possible function that can be performed on
any parallel process --- the one that renders a given pixel in the output image.
Since control-flow analysis is able to determine that, the runtime's scheduler
libraries themselves become specialized via copy propagation to jump directly to
that function, rather than relying on an indirect jump through the work-stealing
structures.
This transformation also frees up many other variables that were kept by the
scheduler to track the work in progress.
While this program is certainly a special case, it is a good example of the
usefulness of these optimizations in a whole-program compiler, as it allows
specializations that are not available to the application author.

The only statistically significant negative performance impact is on the
one-processor version of Quicksort.
This benchmark, unfortunately, shows one of the risks of these optimizations.
Copy propagation and inlining can potentially extend the live range of
variables.
In the one processor version of this benchmark, the live range of one variable
that would otherwise have been copied into a closure once and forgotten is
extended into a another function that did not previously reference it.
This extension not only increases the size of that function's closure, but also
requires the value be captured many times.
In cases with more processors, the other optimizations balance out this one bad
case, but it does demonstrate one of the risks of inlining or performing copy
propagation on functions with free variables.

\begin{table*}[ht]
\begin{center}
\begin{tabular}{r || r | r | r || r | r | r || r | r | r}
\multicolumn{1}{c||}{} & \multicolumn{3}{c||}{1 Processor} & \multicolumn{3}{c||}{16 Processors} & \multicolumn{3}{c}{Optimizations} \\
 &  &  &  & & &  & Branch & Copy &   \\
Benchmark & Speedup & Median & Std. Dev. & Speedup & Median & Std. Dev. & \multicolumn{1}{c|}{Elim.} & Prop. & Inlined  \\
\hline
Barnes-hut & 0\% & 17.2 & 0.03 & -1.6\% & 2.1 & 0.048 & 3 & 14 & 0 \\
Mandelbrot & 12\% & 7.97 & 0.019 & 7.8\% & 0.646 & 0.0445 & 1 & 3 & 0 \\
Quickhull & 0.8\% & 2.86 & 0.003 & 1.3\% & 0.232 & 0.012 & 3 & 14 & 0 \\
Quicksort & -3.3\% & 14.05 & 0.02 & 0.2\% & 1.18 & 0.025 & 1 & 5 & 0 \\
Raytracer & 0.1\% & 15.26 & 0.01 & 4.4\% & 1.160 & 0.074 & 1 & 3 & 0 \\
SMVM & 0.3\% & 6.175 & 0.005 & -0.6\% & 0.669 & 0.019 & 2 & 8 & 5 \\
\end{tabular}%
\end{center}
\caption{
  Performance results from branch elimination, copy propagation, and higher-order inlining optimizations.
  Medians and standard deviation are for the optimized version, reported in seconds. 
}
\label{benchmarks:optimization}
\end{table*}%


%
\section{Related Work}
\label{sec:related}
The problem of detecting when two environments are the same with respect to some
variables is not new.
It was first given the name \emph{environment consonance} in Shivers'
Ph.D. thesis~\cite{shivers:phd-thesis}.
He proposed checking this property by re-running control-flow analysis (CFA)
incrementally --- at cost polynomial in the program size --- at each inlining
point.

Might revisited the problem in the context of his Ph.D. thesis, and showed
another form of analysis, \dcfa{}, which more explicitly tracks environment
representations and can check for safety without re-running the
analysis at each inlining point~\cite{might:delta}.
Unfortunately, this approach also only works in theory, as while its runtime is
faster in practice than a full 1CFA (which is exponential), it is not scalable
to large program intermediate representations.
Might also worked on anodization, which is a more recent technique that
identifies when a binding will only take on a single value, opening up the
possibility of several optimizations similar to this one~\cite{might:shape}. 

Reps, Horowitz, and Sagiv were among the first to apply graph reachability to
program analysis~\cite{interprocedural-analysis}, focusing on dataflow and
spawning an entire field of program analyses for a variety of problems, such as
pointer analysis and security.
While they also present an algorithm for faster graph reachability, theirs is
still polynomial time, which is far too slow for the number of nodes in our
graphs.
A different algorithm for graph reachability that has even better asymptotic
performance than the one we present in \secref{sec:reachability} is also
available~\cite{efficient-closure}, computing reachability at the same time that
it computes the strongly-connected components.
However, it relies on fast language implementation support for mutation, which
is not the case in our compiler's host implementation system, Standard ML of New
Jersey~\cite{smlnj}, so we use an algorithm that better supports the use of
functional data structures.

Serrano's use of 0CFA in the Bigloo compiler is the most similar to our work
here~\cite{serrano:cfa-paradigm}.
It is not discussed in this paper, but we similarly use the results of CFA to
optimize our closure generation.
In that paper, he does not discuss the need to track function identifiers within
data types (\eg{}, lists in Scheme) or limit the depth of that tracking, both of
which we have found crucial in ML programs where functions often are at least in
tuples, due to the default calling convention.
Bigloo does not perform inlining of functions with free variables.

Waddell and Dybvig use a significantly more interesting inlining heuristic in
Chez Scheme, taking into account the potential impact of other optimizations to
reduce the size of the resulting code, rather than just using a fixed threshold,
as we do~\cite{chez-inlining}.
While they also will inline functions with free variables, they will only do so
when either those variables can be eliminated or they know the binding at
analysis time.
Our approach differs from theirs in that we do not need to know the binding at
analysis time and we support whole-program analysis, including all referenced
library functions.


%
\section{Conclusion}
\label{sec:concl}
In this work, we have demonstrated the first \emph{practical} and general
approach to inlining and copy propagation that reasons about the environment.
We hope that this work ushers in new interest and experimentation in
environment-aware optimizations for higher-order languages.
Additionally, we have presented tuning techniques for a control-flow analysis
that tracks datatypes and an optimization --- branch elimination --- that is
both useful and nearly free in a compiler that already performs control-flow
analysis. 

\subsection{Future work}
We have not investigated other optimizations, such as rematerialization, that
were presented in some of Might's recent work on anodization~\cite{might:shape}
and might have an analog in our framework.
An obvious extension to branch elimination is \lstinline{case} statement
specialization based on the flow of datatypes through the program.
In this optimization, we could take an ML function that pattern matches across a
datatype's constructors and remove the arms of the pattern match that we can
statically guarantee will never flow to that function.

Additionally, our control-flow analysis needs further optimizations, both to
improve its runtime and its precision.
We have previously investigated Hudak's work on abstract reference
counting, which resulted in improvements in both runtime and
precision,\footnote{Best results were achieved when using a \lstinline{maxrc} of
  1.} but that
implementation is not yet mature~\cite{abstract-ref-counting}.

\acks
David MacQueen, Matt Might, and David Van Horn all spent many hours discussing
this problem with us, and without their valuable insights this work would likely
have languished.

This material is based upon work supported by the National Science Foundation
under Grants CCF-0811389 and CCF-1010568, and upon work performed in part while
John Reppy was serving at the National Science Foundation.
The views and conclusions contained herein are those of the authors and should
not be interpreted as necessarily representing the official policies or
endorsements, either expressed or implied, of these organizations or the
U.S.\ Government.

\bibliographystyle{common/alpha}
\bibliography{paper.bbl}

\end{document}